\documentclass[conference]{IEEEtran}
\usepackage{ngerman}
\usepackage[utf8]{inputenc}
\usepackage{cite}
\usepackage[printonlyused,nolist]{acronym}
\usepackage{enumitem}
\usepackage[textwidth=30]{todonotes}
\usepackage{booktabs}
\usepackage[hidelinks]{hyperref}
\usepackage{color}

\usepackage{subfig}

\begin{acronym}
  \acro{DDoS}{Distributed Denial of Service}
  \acro{C2}{Command and Control}
  \acro{P2P}{Peer-to-peer}
  \acro{MM}{Membership Management}
  \acro{NL}{Neighborlist}
  \acro{BCS}{Bogus Command Sequence}
  \acro{SPoF}{Single Poing of Failure}
  \acro{GDPR}{General Data Protection Regulation}
\end{acronym}

\begin{document}
\title{I Trust my Zombies: A Trust-enabled Botnet}

\author{
	\IEEEauthorblockN{Emmanouil Vasilomanolakis\IEEEauthorrefmark{1}, Jan Helge Wolf\IEEEauthorrefmark{1}, Leon Böck\IEEEauthorrefmark{1}, \\Shankar Karuppayah\IEEEauthorrefmark{2}, Max Mühlhäuser\IEEEauthorrefmark{1}}
	
	\IEEEauthorblockA{\IEEEauthorrefmark{1}
		Telecooperation Lab,
		Technische Universität Darmstadt\\ Darmstadt, Germany \\
		\{vasilomano, boeck, max\}@tk.tu-darmstadt.de,  \{janhelge.wolf\}@stud.tu-darmstadt.de
	}
	
	\IEEEauthorblockA{\IEEEauthorrefmark{2}
		National Advanced IPv6 Centre,
		Universiti Sains Malaysia, \\Penang, Malaysia	\\
		\{kshankar\}@usm.my
	}
}

\maketitle

\begin{abstract}
Defending against botnets has always been a cat and mouse game. Cyber-security researchers and government agencies attempt to detect and take down botnets by playing the role of the cat. In this context, a lot of work has been done towards reverse engineering certain variants of malware families as well as understanding the network protocols of botnets to identify their weaknesses (if any) and exploit them. While this is necessary, such an approach offers the botmasters the ability to quickly counteract the defenders by simply performing small changes in their arsenals. 

We attempt a different approach by actually taking the role of the Botmaster, to eventually anticipate his behavior. That said, in this paper, we present a novel computational trust mechanism for fully distributed botnets that allows for a resilient and stealthy management of the infected machines (zombies). We exploit the highly researched area of computational trust to create an autonomous mechanism that ensures the avoidance of common botnet tracking mechanisms such as sensors and crawlers. In our futuristic botnet, zombies are both smart and cautious. They are cautious in the sense that they are careful with whom they communicate with. Moreover, they are smart enough to learn from their experiences and infer whether their fellow zombies are indeed who they claim to be and not government agencies' spies. We study different computational trust models, mainly based on Bayesian inference, to evaluate their advantages and disadvantages in the context of a distributed botnet. Furthermore, we show, via our experimental results, that our approach is significantly stronger than any technique that has been seen in botnets to date.
\end{abstract}

\section{Introduction}

Botnets are networks of infected computing devices, called bots. These bots can be remotely controlled and instructed to conduct criminal activities by malicious entities that are commonly referred to as botmasters. Botnets are used for a multitude of malicious activities such as \ac{DDoS}, banking theft or spam email distribution. 
For this reason, researchers attempt to defend against botnets by proposing novel detection and prevention methods; for instance, intrusion detection systems, honeypots, etc. \cite{vasilomanolakis2015survey,provos2007}. 

Traditionally, many botnets have been based on a centralized architecture consisting of a \ac{C2} server that relays commands directly to the bots. However, this architecture presents a \ac{SPoF} in the centralized server which can be used to seize control of the botnet. Therefore, more advanced botnets implement a \ac{C2} channel based on unstructured \ac{P2P} overlays. These botnets do not inherit the \ac{SPoF} of centralized approaches. Furthermore, they are very resilient to node churn and node removal attacks \cite{rossow2013p2pwned}.

As the lack of a central server prevents easy monitoring, researchers have developed various means for gathering intelligence in \ac{P2P} botnets. This is usually achieved by first reverse engineering the communication protocol and afterwards deploying crawlers and sensors to enumerate the botnet population. Nevertheless, botnets such as Sality \cite{falliere2011b} or GameOver Zeus \cite{andriesse2013} already implement features to impede monitoring attempts.

Within this work we present a novel approach to thwart monitoring attempts by researchers and law-enforcement agencies. The proposed mechanism is based on the utilization of computational trust along with special crafted messages that the bots exchange to verify the correct behavior of their peers. Our work is one among others published recently that present means to detect monitoring operations in \ac{P2P} botnets \cite{Bock2015, Karuppayah2016, karuppayah2017sensorbuster, Andriesse2015}. This suggests that the options to harden \ac{P2P} botnets are manifold and may eventually prevent successful monitoring entirely. Therefore, we want to highlight that the need for developing new mechanisms to efficiently gather intelligence on \ac{P2P} botnets is urgent.

\begin{figure*}[htp!] 
	\centering
	\subfloat[Visualization of sensor popularity after joining the network]{%
		\includegraphics[width=0.5\textwidth]{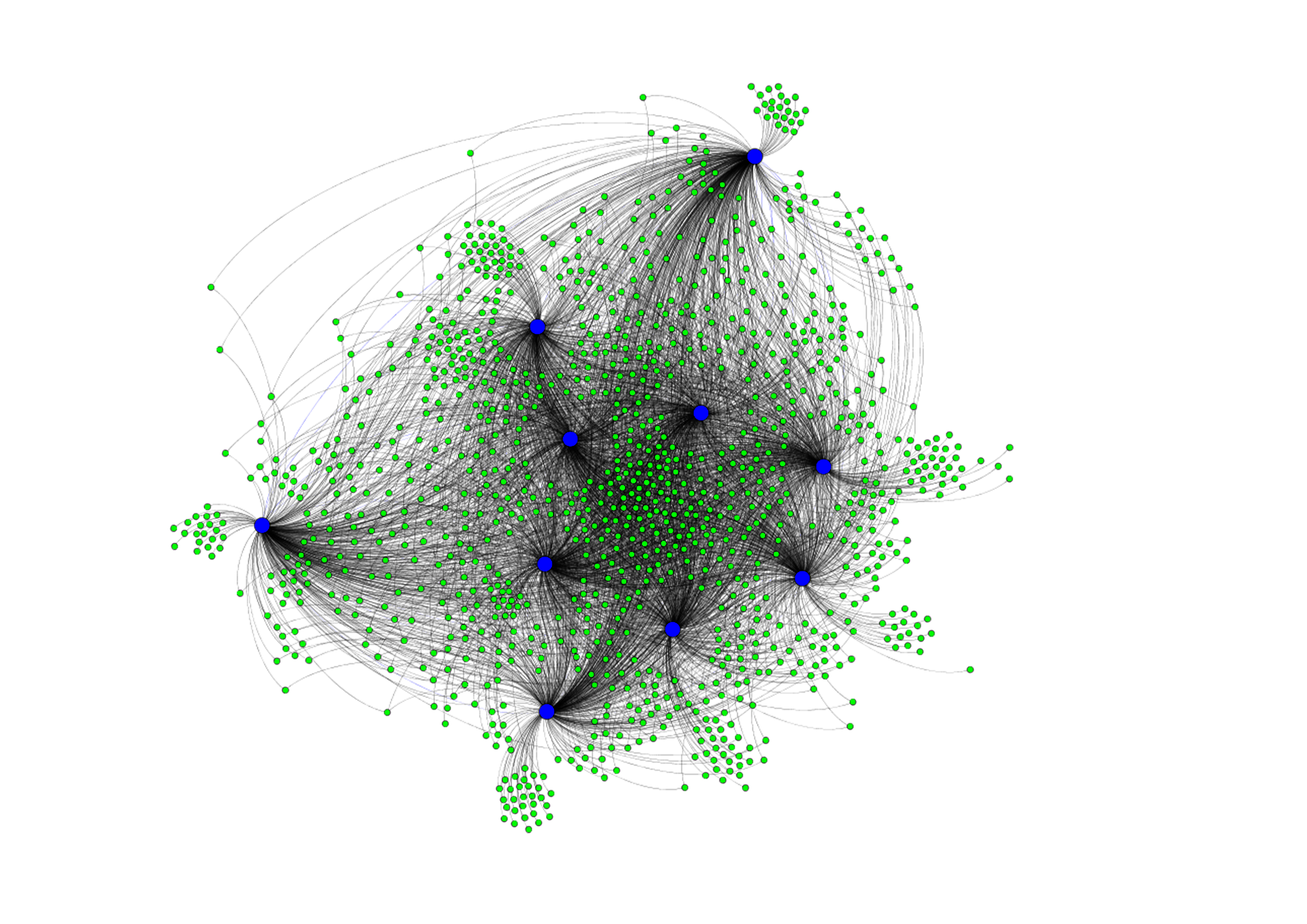}}
	\label{1a}\hfill
	\subfloat[Visualization of sensor popularity $14$ days later.]{%
		\includegraphics[width=0.5\textwidth]{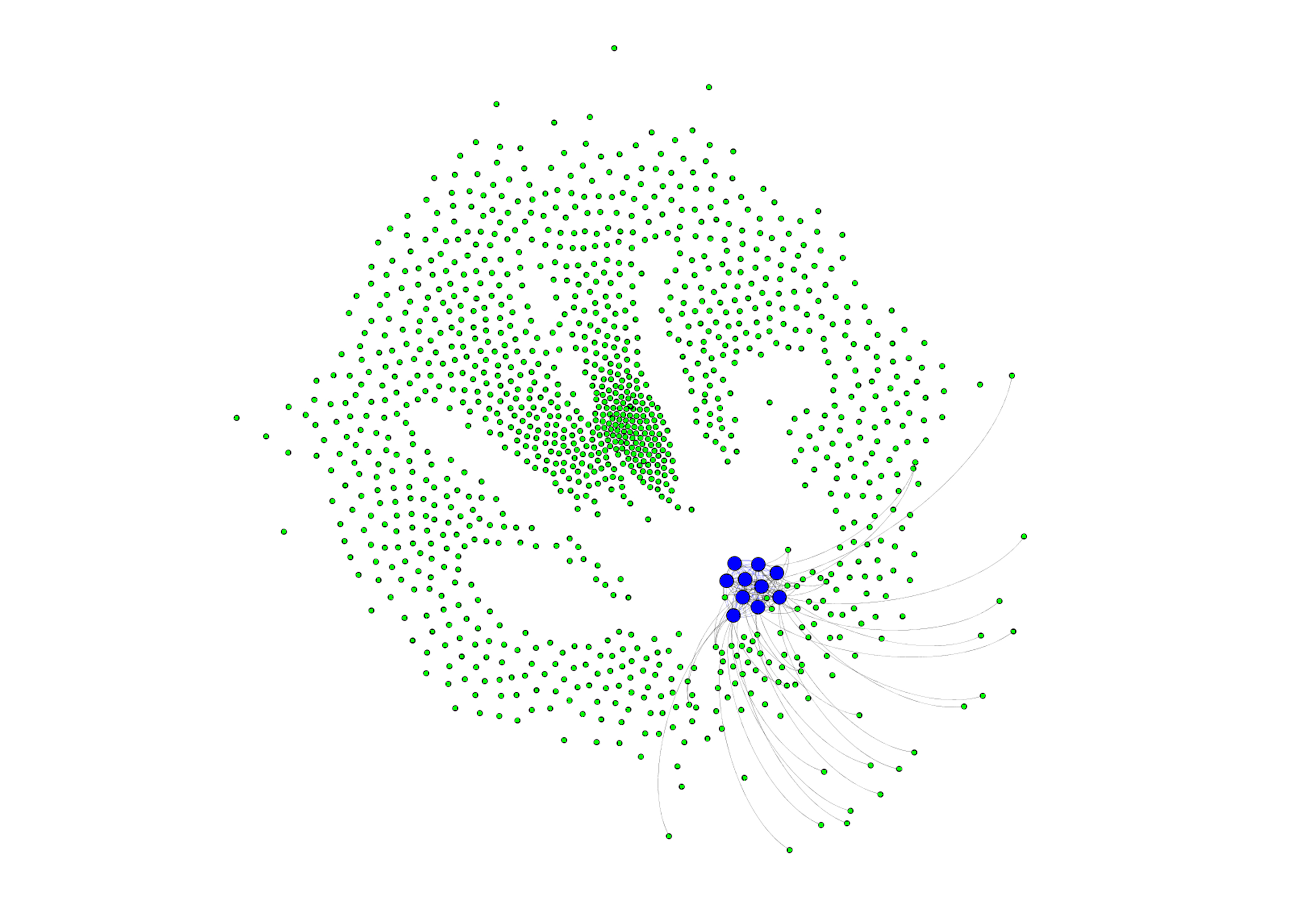}}
	\label{1b}\hfill
	\caption{Visualization of sensor popularity after joining the network and $14$ days later. Sensors are depicted in blue and benign bots in green.}
	\label{fig1} 
\end{figure*}

\section{Trust Enabled Sensor Detection on \ac{P2P} Botnets}
\label{sec:methods}
In the following, we will introduce our trust mechanism based approach to detect sensors in \ac{P2P} botnets. For this, we first introduce some background on \ac{P2P} botnet \ac{MM} and computational trust mechanisms. Afterwards, we explain how computational trust can be used to identify and automatically blacklist sensor nodes deployed by researchers or law enforcement agencies.

\paragraph{Botnet Membership Management}
\label{par:mm}

To ensure that the \ac{P2P} botnet remains connected in the presence of churn, i.e., nodes joining and leaving the network, a \ac{MM} system is used to frequently update connection information. Each bot in a \ac{P2P} network maintains a list of other bots. This list is commonly referred to as \ac{NL} and the bots stored within the \ac{NL} are called neighbors. 

Each bot regularly contacts its neighbors to check their responsiveness as well as to receive updated commands. If all neighbors are unavailable, a bot is isolated from the botnet and will not be able to receive any updates or botmaster commands. Therefore, it is important to update the \ac{NL} frequently by replacing inactive neighbors with other active bots. This is accomplished by sending probing messages to all bots in the \ac{NL} for availability within a fixed interval, called \ac{MM} cycle. These probing messages are commonly referred to as \textit{hello} messages. 

If a node remains unresponsive for a prolonged period of time, it will be replaced by a ``fresh'' entry of an online bot. Furthermore, botnets also use the \ac{MM} cycle to exchange information about the ID of the latest instruction set. If one bot does not have the most current update it will query a neighbor to forward the latest instruction set. In the case of the Sality botnet, this ID is directly embedded in the \textit{hello} and \textit{hello reply} messages.

\paragraph{Computational Trust}
Trust is a familiar term for human beings, who make a plethora of trust-based decisions on a daily basis. Usually a person or \textit{trustor} engages with another person or \textit{trustee} based on the assumption that the trustee will behave as the trustor expects it \cite{marsh1994formalising}. 

Computational trust provides a means to model the concept of trust for computers and other devices. In particular, evidence-based trust mechanisms use experiences, collected from past interactions, to predict the future behavior of the trustee. These experiences can either be first-hand experiences from prior interactions with the trustee or second-hand experiences shared by other trustors through recommendations or referrals \cite{artz2007survey,ries2009trust}.

\paragraph{Disclosing Sensor Nodes}
Our mechanism, for disclosing sensor nodes, builds on the assumption, that sensors and crawlers will not aid the botnet in any way or form. This is a common assumption \cite{karuppayah2017sensorbuster}, as even law-enforcement agencies have to adhere to national and international laws as well as ethics. 
In more details, our work assumes that a sensor is not allowed to participate in malicious activities of the botnet and/or to disseminate (to other benign bots) new versions of malware or command sets originating from the botmaster.

We exploit this, by introducing a new type of message, called \ac{BCS} message, which is designed to disclose the unwillingness of sensors to participate in criminal activities\footnote{Note that our work can be easily extended to crawlers as well. However, since crawlers are relatively easy to detect, we consider them out of the scope of this paper.}. As we have explained in Section \ref{par:mm}, bots frequently exchange \textit{hello} messages that include the ID of the latest botmaster command. In the \ac{BCS} message a bot does not attach its real command \textit{ID}, but instead chooses a significantly lower value. A regular bot will respond to this message with its current command ID and the latest update attached in the reply. However, a sensor cannot forward a valid command update without violating the assumption that it may not participate in criminal activities.

A bot will frequently send these \ac{BCS} messages to its neighbors to probe them for their trustworthiness. Upon receiving anything but a recent command ID together with a valid command update, the engagement will be considered a negative experience. We use the recorded experiences together with evidence-based trust models to make trust-based decisions.
%blacklisting explanation
In more details, when the calculated trust score, of a bot, falls below a certain (predefined) threshold, this bot is considered to be a sensor. As such, the sensor is removed from the \ac{NL} of the bot, added to a blacklist and all incoming messages in the future will be ignored.

To avoid engaging in criminal activities, a sensor could respond to a \ac{BCS} message in three different ways: 
\begin{itemize}
	\item it can reply with the same command \textit{ID} 
	\item  it may not reply at all
	\item it can attempt to corrupt the payload on purpose before sending a reply
\end{itemize}

Note that while each of these replies is considered a negative experience by a bot, it is possible that a real bot responds similarly on rare occasions. As an example, a response may actually be corrupted due to certain network problems, or a bot might go offline during the interaction (and therefore it does not send a response).

To avoid blacklisting a bot preemptively based on a single negative experience, we record multiple experiences. These are then used as an input to make blacklisting decisions given a computational trust model. To identify which model is best suited, we evaluated our approach using four different computational trust models. Namely these are, the \textit{ebay} user rating trust model, the \textit{beta distribution}, \textit{subjective logic} \cite{josang2001logic}, and \textit{certain trust} \cite{ries2009trust}. Our preliminary results indicate, that the ebay trust model performs the best, even though it is the most basic of all four models\footnote{It should be noted, however, that the \textit{ebay} system, being the simplest one, is the only one who introduces (a low number) of false positives. In contrast, the remaining three computational trust mechanisms can achieve a precision of $1$.}. 

In Figure \ref{fig1}, the connectivity of $10$ sensors is depicted at the beginning of a simulation and after $14$ days of simulation. As it is depicted in the figure, the popularity, i.e. the in-degree, of all sensors decreases significantly throughout the simulation. In fact, with the ebay trust model we were able to reduce the popularity of sensors by more than $97\%$ in comparison to the original Sality botnet protocol.

\section{Conclusion \& Future Work}

We have shown, that computational trust can be used as a mechanism to greatly diminish the monitoring information that can be obtained with sensor nodes. To the best of our knowledge, this is the first work that approaches anti-monitoring mechanisms from a non graph-theoretic perspective. Heretofore, the state of the art in anti-monitoring mechanisms has been utilizing protocol-level anomaly detection \cite{Andriesse2015} and/or graph-theoretic approaches to detect the activity of crawlers or sensors in \ac{P2P} botnets \cite{Bock2015, karuppayah2017sensorbuster}. 

We argue that the work presented here is one of many different possible anti-monitoring mechanisms that can be deployed in the \ac{P2P} botnets of the near future. Therefore, we want to press the issue that \ac{P2P} botnet monitoring will not be possible to the same extent as it is now. Furthermore, legal and ethical boundaries greatly restrict the range of options for researchers and law-enforcement. In fact, such limitations are expected to further grow in the future; for instance via the enforcement of the European Union's \ac{GDPR} \cite{green2017ransomware}.
Finally, we argue that collaborative monitoring may be a way to mitigate the effect of some anti-monitoring mechanisms. Nevertheless, due to the sheer amount of possible anti-monitoring mechanisms, we strongly believe that regulators and researchers have to work together to develop botnet monitoring mechanisms that can not be easily detected by botmasters while adhering to the applicable legal systems.

% %future work
In our future work, we plan to present our computational trust-based method in a more formal and detailed manner. In addition, we are currently performing full-fledged simulations to measure the extent of our method's performance in a highly realistic scenario. Moreover, we plan to further analyze the usage of colluding sensors and their effectiveness, for collaborative monitoring, in such a resilient botnet environment.

\bibliographystyle{plain}
\bibliography{bib/Cybersecurity-Botnets,bib/lit,bib/manual,bib/Cybersecurity}

\end{document}